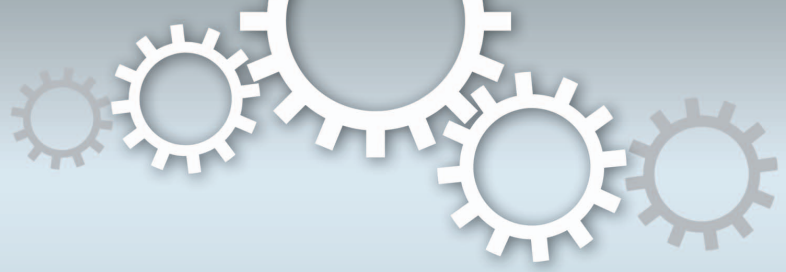

# SCIENTIFIC REPORTS

**OPEN**

## Latent instabilities in metallic LaNiO$_3$ films by strain control of Fermi-surface topology

Hyang Keun Yoo[1,2], Seung Ill Hyun[3], Luca Moreschini[4], Hyeong-Do Kim[1,2], Young Jun Chang[4,5], Chang Hee Sohn[1,2], Da Woon Jeong[1,2], Soobin Sinn[1,2], Yong Su Kim[1,2], Aaron Bostwick[4], Eli Rotenberg[4], Ji Hoon Shim[3,6] & Tae Won Noh[1,2]

[1]Center for Correlated Electron Systems, Institute for Basic Science (IBS), Seoul 151-747, Republic of Korea, [2]Department of Physics and Astronomy, Seoul National University, Seoul 151-747, Republic of Korea, [3]Department of Chemistry, Pohang University of Science and Technology, Pohang 790-784, Republic of Korea, [4]Advanced Light Source (ALS), E. O. Lawrence Berkeley National Laboratory, Berkeley, California 94720, USA, [5]Department of Physics, University of Seoul, Seoul 130-743, Republic of Korea, [6]Division of Advanced Nuclear Engineering, Pohang University of Science and Technology, Pohang 790-784, Republic of Korea.

Strain control is one of the most promising avenues to search for new emergent phenomena in transition-metal-oxide films. Here, we investigate the strain-induced changes of electronic structures in strongly correlated LaNiO$_3$ (LNO) films, using angle-resolved photoemission spectroscopy and the dynamical mean-field theory. The strongly renormalized $e_g$-orbital bands are systematically rearranged by misfit strain to change its fermiology. As tensile strain increases, the hole pocket centered at the $A$ point elongates along the $k_z$-axis and seems to become open, thus changing Fermi-surface (FS) topology from three- to quasi-two-dimensional. Concomitantly, the FS shape becomes flattened to enhance FS nesting. A FS superstructure with $Q_1 = (1/2,1/2,1/2)$ appears in all LNO films, while a tensile-strained LNO film has an additional $Q_2 = (1/4,1/4,1/4)$ modulation, indicating that some instabilities are present in metallic LNO films. Charge disproportionation and spin-density-wave fluctuations observed in other nickelates might be their most probable origins.

Misfit strain has been a key control parameter to tune physical properties and to search for new emergent phenomena in transition-metal-oxide (TMO) films[1–9]. Recently, rare-earth nickelate films, $R$NiO$_3$ ($R$ = rare-earth element), have attracted much interest to control their metal-insulator transitions or peculiar 'up-up-down-down' spin orderings by misfit strain[10–13]. Among them, LaNiO$_3$ (LNO) was considered to be less interesting, because its bulk is a paramagnetic metal at all temperatures[14]. However, after theoretical proposals that it could become an electronic analogue to high-$T_c$ cuprates by strain and layer-thickness control[15–17], a lot of studies have been devoted to realize it[18–28]. Though its final goal has not been achieved yet, the LNO hetero-structures revealed various intriguing properties such as metal-insulator or antiferromagnetic (AFM) transitions by dimensionality control[21–27] and a signature of charge disproportionation (CD), absent in bulk, by strain control[28].

The CD in $R$NiO$_3$, except for La, is ascribed to very small or negative charge-transfer energy $\Delta$ from O 2$p$ to Ni 3$d$ orbitals[29,30]. In LNO films, the misfit strain will induce a change of Madelung potential, resulting in a different $\Delta$[28], which would be an origin of CD in LNO films[28,30]. Chakhalian et al. predicted that the CD may occur under tensile strain[28], but Lau and Millis proposed a phase diagram in which a charge ordering can occur at all strain conditions[30]. Meanwhile, recent theoretical calculations predicted that a spin-density-wave (SDW) order can emerge in metallic LNO thin films[30–32]. Especially, Lee et al. claimed that the SDW order should appear due to Fermi-surface (FS) nesting by tuning effective hopping parameters between Ni 3$d$ orbitals[31,32]. In spite of these intriguing theoretical predictions, there have been few experimental evidences for them.

In this Letter, to investigate the systematic changes of electronic structures of LNO films under various misfit strains, we employed *in situ* angle-resolved photoemission spectroscopy (ARPES) and the dynamical mean-field theory (DMFT). As the strain states change from compressive to tensile, we observed that the $e_g$-band rearrangement systematically changes the fermiology of LNO films. Additionally, all the LNO films show a FS super-structure with $Q_1 \equiv (1/2,1/2,1/2)$, implying an existence of CD fluctuations. A tensile-strained LNO film shows another FS superstructure with $Q_2 \equiv (1/4,1/4,1/4)$, suggesting an existence of a SDW instabilities due to significant FS nesting in the quasi-two-dimensional (2D) FS. Our results demonstrate how such latent instabilities are affected by misfit strain in LNO films. Thus, misfit strain can be used as a tuning parameter to search for latent orders in TMO films.

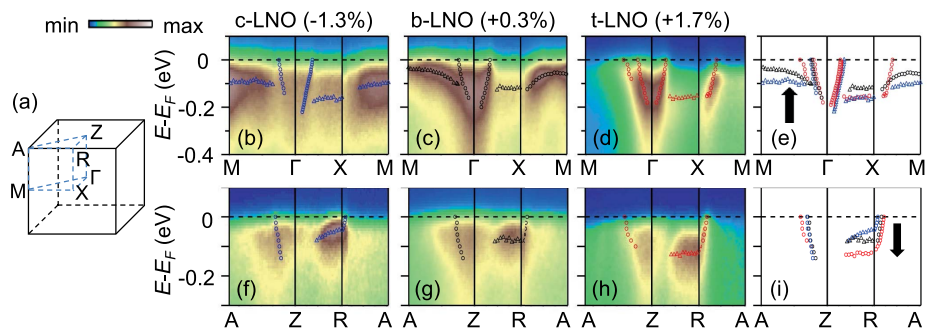

**Figure 1 | Strain control of the electronic band dispersions in LNO films.** (a) BZ of tetragonal LNO. ARPES spectra along the $M\Gamma XM$ and the $AZRA$ lines of (b,f) $c$-LNO, (c,g) $b$-LNO, and (d,h) $t$-LNO, respectively. Quasiparticle-peak positions from MDCs and EDCs are represented by empty circles and triangles, respectively. Overlay of quasiparticle-peak positions under the different strain states along (e) the $M\Gamma XM$ and (i) the $AZRA$ lines.

## Results

**Orbital-dependent electronic-structure changes induced by misfit strain.** High-quality LNO films were epitaxially grown using pulsed laser deposition (PLD) on (001)-oriented $LaAlO_3$ (LAO), $NdGaO_3$ (NGO), and $SrTiO_3$ (STO) substrates. Due to lattice mismatches, the LNO films on LAO, NGO and STO substrates should be under compressive strain ($-1.3\%$, hereafter denoted by $c$-LNO), nearly free strain, i.e. bulk-like ($+0.3\%$, $b$-LNO), and tensile strain ($+1.7\%$, $t$-LNO), respectively. Since the LNO films have tetragonal distortions due to biaxial strain, we will use, in this paper, Brillouin zone (BZ) notation of the tetragonal structure for the symmetric points, as shown in Fig. 1a. Bulk LNO has a simple three-dimensional (3D) electronic structure composed of a small electron pocket centered at the $\Gamma$ point and a large hole pocket at the $A$ point[33,34].

Figures 1b–d show ARPES spectra of $c$-LNO, $b$-LNO, and $t$-LNO along the $M\Gamma XM$ line. Quasiparticle-peak positions from intensity maxima in momentum or energy distribution curves (MDCs or EDCs) are also plotted by empty circles and triangles, respectively. Figure 1e shows an overlay of the quasiparticle-peak positions of the three LNO films, which clearly shows an upward shift of the bands as the misfit strain changes from compressive to tensile. Similar plots are presented in Figs. 1f–i along the $AZRA$ line. As seen in Fig. 1i, in contrast to the $M\Gamma XM$ line, the band makes a downward shift.

To understand the strain-dependent band shifts, we compared our ARPES data (blue triangles) with the results of generalized gradient approximation (GGA, solid lines) and GGA + DMFT (orange-scale images) calculations in Figs. 2a–d (see Supplementary Information (SI) for calculation methods). Note that misfit-strain values in the calculations are a little different from experimental ones, since we chose the values at which converged solutions were obtained. See the Fig. 2 caption. On the other hand, the experimental $k_z$ resolution due to short photoelectron escape depth is about 0.13 Å$^{-1}$ (full width at half maximum), so the GGA + DMFT results were presented after averaging over the $k_z$ resolution in the whole figures (see SI for a detail). As clearly seen in the figure, the GGA + DMFT describes the band dispersions much better than the GGA, which implies strong electron-electron correlations in all LNO films[35,36]. According to the calculations, the hole band along the $M\Gamma XM$ line is mostly composed of a $d_{3z^2-r^2}$ orbital (SI Fig. S6). Under tensile strain, the smaller the distance between the $d_{3z^2-r^2}$ orbital and apical oxygen ions becomes, the stronger the hybridization. Thus, it pushes the anti-bonding $d_{3z^2-r^2}$ band toward higher energy as in Figs. 2a and 2c. On the other hand, the hole band along the $AZRA$ line is mostly composed of a $d_{x^2-y^2}$ orbital, thus we can expect opposite behavior as seen in Figs. 2b and 2d.

**Strain control of Fermi-surface topology.** Comparing the hole-band dispersions in Figs. 1b–d, we can find a hint that the FSs of the LNO films can be changed significantly due to misfit strain. While the bands of the $c$- and the $b$-LNO's are located below the Fermi level ($E_F$), for $t$-LNO, the bands cross the $E_F$ to modify a FS shape. Because of the $k_z$ resolution, we can suspect that the bands may not cross the $E_F$ in a real situation. However, as seen in Fig. 3b, the hole pocket size in the $\Gamma XM$ plane is larger than that from the GGA + DMFT. Thus, if the fermiology changes smoothly, we can expect that the hole pocket is more open along the $k_z$-axis than the calculated one in Fig. 3f. Such a change in $t$-LNO cannot be explained simply by the GGA, in which the upward level shift of the hole band by the misfit strain is only about 0.2 eV (Figs. 2a and 2c). On the other hand, the GGA + DMFT results indicate that strong electron correlations push the hole band further in $t$-LNO. As a result, the bands make contact with the $E_F$ along the $M\Gamma$ and the $XM$ lines. This result suggests that both misfit strain and strong correlations should play an important role in the fermiology change in LNO films.

Figures 3a and 3b show this FS change (left panel: ARPES; right panel: GGA + DMFT) in the $\Gamma XM$ plane. In $c$-LNO, there is only the electron pocket at the $\Gamma$ point, whereas in $t$-LNO we can see a hole pocket centered at the $M$ point. The intense feature at the $M$ point in $c$-LNO does not originate from $E_F$-crossing bands but from the $k_z$ resolution and the broadening of the hole band near the $E_F$. This fermiology change in the $\Gamma XM$ plane is well described by the GGA + DMFT, though the size of the hole pocket in $t$-LNO is smaller than the experimental one. FS maps in the $ZRA$ plane also reveal another interesting FS change, as shown in Figs. 3c and 3d. As the strain state goes from compressive to tensile, the hole-pocket shape becomes a nearly perfect square with a large flat region, thus enhancing FS nesting. The details of the FS nesting and its possible relationship to density waves will be discussed later.

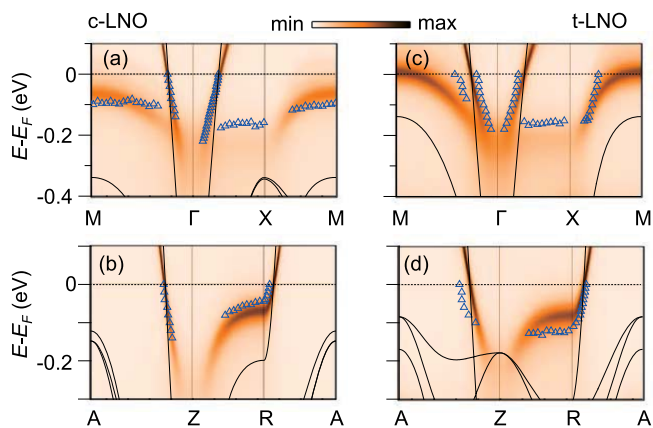

**Figure 2 | Calculated band dispersions with respect to misfit strain.** LNO band dispersions from GGA (solid lines) and GGA + DMFT (orange-scale images) calculations along the $M\Gamma XM$ and the $AZRA$ lines calculated with (a,b) $-1.7\%$ compressive ($c$-LNO) and (c,d) $+2.0\%$ tensile strain ($t$-LNO), respectively. The experimental band dispersions are displayed by empty blue triangles.

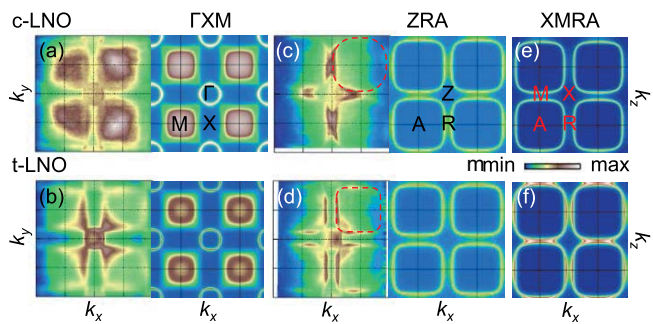

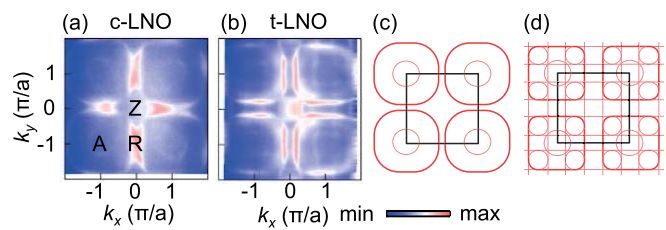

**Figure 3 | Fermiology change of LNO films in response to misfit strain.** FS maps in the $\Gamma XM$ and the $ZRA$ planes of (a,c) $c$-LNO and (b,d) $t$-LNO, respectively. Left (Right) panels are for ARPES (GGA + DMFT). The red dashed lines in (c) and (d) are guidelines for better comparison of the FS shapes. FS maps in the $XRAM$ plane of $c$-LNO and $t$-LNO are presented in (e) and (f) from GGA + DMFT, respectively.

To figure out the whole FS shape, FS maps in the $XRAM$ plane from the GGA + DMFT calculations are presented in Figs. 3e and 3f. For $c$-LNO, two hole pockets centered at the $A$ points are well separated. On the other hand, for $t$-LNO, the two hole pockets merge together around the $M$ point, resulting in an open FS along the $k_z$-axis. Thus, the misfit strain can drive a Lifshitz transition, i.e. the hole FS topology evolves from 3D to quasi-2D as the strain state changes from compressive to tensile.

**Two kinds of Fermi-surface superstructures depending on misfit strain.** A careful examination of the FS maps in Figs. 3c and 3d reveals faint circles and lines in the $ZRA$ plane, implying the existence of FS superstructures. Such superstructures are often related with FS instabilities, which can cause various intriguing emergent phenomena such as charge or spin density waves[37,38] and superconductivity[39]. Thus, proper understanding of the FS superstructures will provide us valuable insights on possible emergent phenomena.

To investigate the FS superstructures more clearly, we constructed constant-energy-surface (CES) maps in the $ZRA$ plane at a binding energy of 30 meV using a log scale, as displayed in Figs. 4a and 4b for $c$- and $t$-LNO, respectively. In $c$-LNO, we can observe a circular-shape superstructure at the $A$ point, which is similar to that observed at the $\Gamma$ point, as shown in Fig. 3a. Thus, this superstructure can be considered as a replica of the electron pocket. To reproduce this superstructure, we folded the whole BZ with $\mathbf{Q}_1$ and plotted a resulting CES map in Fig. 4c. This schematic CES map represents the experimental map quite well, suggesting that a $\mathbf{Q}_1$ modulation should exist in $c$-LNO. Note that the FS superstructure in the $\Gamma XM$ plane is also consistent with the $\mathbf{Q}_1$ modulation and that $b$-LNO also shows a similar superstructure (SI Figs. S9 and S10).

On the other hand, the CES map for $t$-LNO displays more complicated superstructures, as shown in Fig. 4b. In addition to the superstructure at the $A$ point, it shows additional one centered at the $Z$ point with a rounded-square shape. To reproduce this super-structure, we folded the hole pocket in the BZ with $\mathbf{Q}_2$ in addition to the folding of the whole BZ with $\mathbf{Q}_1$, which results in a CES map shown in Fig. 4d. This schematic CES map matches well the complicated superstructure in Fig. 4b. This result suggests that $t$-LNO should have another periodic modulation with $\mathbf{Q}_2$ for the hole pockets. Note that the periodic modulation with $\mathbf{Q} = (1/4,1/4,0)$ will give a similar superstructure in Fig. 4b, but it cannot reproduce a FS superstructure in the $\Gamma XM$ plane (SI Fig. S11).

**Figure 4 | Strain-dependent FS superstructures.** Experimental CES maps in the $ZRA$ plane at 30 meV of (a) $c$-LNO and (b) $t$-LNO on a log scale. Schematic CES maps for (c) $c$-LNO after folding the BZ with $\mathbf{Q}_1 = (1/2,1/2,1/2)$ and (d) $t$-LNO after an additional folding of the hole pocket with $\mathbf{Q}_2 = (1/4,1/4,1/4)$.

## Discussion

Now, let us elucidate the origin of the $\mathbf{Q}_1$ modulation. First possibility is a $\sqrt{2} \times \sqrt{2}$ surface reconstruction[40]. Such surface reconstruction will make an electron pocket at the $\Gamma$ point reappear at the $M$ point not at the $A$ point, thus the surface reconstruction cannot be the origin of the $\mathbf{Q}_1$ modulation. Second possibility is the coupling between the misfit strain and metal-oxygen octahedral rotations[41]. Under misfit strain, $NiO_6$ octahedra can be rotated cooperatively, which produces a FS superstructure with a $\mathbf{Q}_1$ modulation. If it occurs, the super-structure intensity in $c$- or $t$-LNO films should be stronger than that in $b$-LNO. However, the superstructure intensity of $b$-LNO is similar to that of $c$-LNO (SI Fig. S10). And we could not observe any significant difference in RHEED patterns during the film growth (SI Fig. S1). Therefore, the observed FS superstructure with the $\mathbf{Q}_1$ modulation in our LNO films should have an electronic origin.

It is well known that 3D $R$NiO$_3$, except $R$ = La, has CD with a rocksalt-pattern lattice distortion[42,43]. Such charge ordering will result in a BZ folding with $\mathbf{Q}_1$. In contrast, a bulk LNO remains as a metal at all temperatures[14], so it is rather difficult to expect such charge ordering in LNO at a first glance. However, recently, a slight CD was predicted to exist even in a metallic bulk $R$NiO$_3$[29,30]. Our observation of the $\mathbf{Q}_1$ modulation supports such theoretical predictions. Note also that ARPES is an ultrafast probe in the femtosecond regime, so it is quite sensitive to short- as well as long-range orders[44,45]. Therefore, the superstructure with the $\mathbf{Q}_1$ modulation could be originated from either short- or long-range CD in the metallic LNO films.

Let us look into the $\mathbf{Q}_2$ modulation. The $R$NiO$_3$, except $R$ = La, has an AFM ordering with an 'up-up-down-down' spin configuration[14,46]. Such magnetic ordering could result in a band folding with $\mathbf{Q}_2$. Since bulk LNO has a paramagnetic ground state[14], such magnetic ordering is not likely to occur. However, Lee et al. recently proposed that if FS nesting in a nickelate becomes strong at $\mathbf{Q}_2$, a SDW order appears even in the metallic regime[31,32]. As we demonstrated earlier with Fig. 3d, the FS of $t$-LNO exhibits a large parallel flat region in the quasi-2D hole pocket. This flattened FS can induce large FS nesting at $\mathbf{Q}_2$. To confirm this intriguing possibility, we calculated susceptibilities from the GGA band dispersions and obtained a sharp peak near $\mathbf{Q}_2$ (SI Fig. S12). This peak in $t$-LNO is sharper than those in other films. This result implies that it is highly plausible that the $t$-LNO film has SDW instabilities with $\mathbf{Q}_2$, consistent with earlier theoretical predictions[31,32]. Actually, recent muon spin rotation[24] and resonant x-ray diffraction[25] measurements showed an AFM ordering in two-unit cell-thick LNO superlattice, but no signature in thicker ones. These results and our observations imply that SDW fluctuations exist even in thicker LNO films. We should note that the nesting vector $Q_2$ is 3D in the quasi-2D FS. This seems to be self-contradictory, but it could be explained by that the spin instability is strongly coupled to the 3D CD.

Appearance of the FS superstructures in our LNO films provides new insights into strain-induced changes of charge and spin orderings in $R$NiO$_3$. Recently, Lau and Millis investigated charge and spin orderings in bulk $R$NiO$_3$ and few-layered $R$NiO$_3$ using the slave-rotor Hartree–Fock formalism[30]. In their proposed phase diagram, there is a 3D metallic region with a slight charge ordering, but no spin ordering in bulk $R$NiO$_3$. Our c-LNO should belong to this 3D metallic region, because it has the 3D electronic structure as in Fig. 3e. And it shows a superstructure with the $Q_1$ modulation only, consistent with a CD instability. On the other hand, in ultrathin $R$NiO$_3$ films, they predicted a 2D metallic region with both charge and spin orderings. Our t-LNO film should belong to this 2D metallic region, because it has the quasi-2D electronic structure with strong FS nesting at $Q_2$ as in Fig. 3f. In addition, its FS map shows superstructures with both the $Q_1$ and $Q_2$ modulations, suggesting coexistence of CD and SDW instabilities. Our study demonstrates that misfit strain can reveal latent instabilities in LNO films.

In conclusion, we systematically investigated how misfit strain can change the electronic structure of LNO films, combining APRES measurements and GGA + DMFT calculations. As the strain state goes from compressive to tensile, strongly renormalized Ni $e_g$ bands are rearranged, and change FS topology, accompanied by flattening of the FS shape in a tensile-strained LNO film. Additionally, in all LNO films, we observed a FS superstructure, which reveals a signature of CD fluctuations. We observed an additional FS superstructure in the tensile-strain case, which can be attributed to SDW fluctuations. Our results suggested that misfit strain can be a useful tuning parameter to search for latent instabilities and have possibilities to induce latent orders in strongly correlated electron systems.

## Methods

**Sample preparation.** High-quality 10 unit-cell-thick LNO films were epitaxially grown using pulsed laser deposition (PLD) on (001)-oriented single-crystalline substrates. To investigate the strain-dependent changes in electronic structures, we grew three kinds of LNO epitaxial films on (001)-oriented LaAlO$_3$ (LAO), NdGaO$_3$ (NGO), and SrTiO$_3$ (STO) substrates. The lattice constants of bulk LNO, LAO, and NGO are 3.84, 3.79, and 3.851 Å in pseudocubic notation, respectively. And the lattice constant of STO is 3.905 Å for the cubic structure. Due to the lattice mismatch, the LNO films on LAO, NGO and STO substrates should be under compressive strain, nearly strain-free, and tensile strain, respectively. For PLD, we used a KrF excimer laser ($\lambda = 248$ nm) with a repetition rate of 2 Hz to ablate sintered stoichiometric targets. The laser energy density was in the range of 1–1.5 J cm$^{-2}$ at the target position. The deposition temperature and the $P(O_2)$ were 600°C and $1 \times 10^{-2}$ Torr, respectively. Then, we cooled down the samples below 100°C in $P(O_2)$, $1 \times 10^{-2}$ Torr, to avoid the oxygen vacancy generation. We monitored the film thickness by using in situ reflection high-energy electron diffraction. We confirmed that all of the LNO films were fully and homogeneously strained by x-ray diffraction at the 9C beamline of Pohang Light Source (PLS). (For the details of film characterizations, see Supplementary Information, Fig. S1).

**In situ ARPES measurements.** ARPES measurements were performed at the Beamline 7.0.1 of ALS, of which end-station was equipped with PLD and ARPES. After deposition, the thin films were transferred to an analysis chamber at a pressure of $5 \times 10^{-11}$ Torr without breaking a vacuum. The sample temperature was kept at 90 K during the measurements. The total energy resolution was 30 meV at $h\nu = 150$ eV. Since LNO films have three-dimensional electronic structures, appropriate photon energies for the high-symmetry planes of the each film were determined by scanning photon energies in the normal emission; the ΓXM (ZRA) plane was found to be located at slightly different energies of 150 eV (114 eV) for LNO/LAO, 154 eV (117 eV) for LNO/NGO, and 157 eV (120 eV) for LNO/STO due to different lattice parameters. For the details of ARPES measurements, see Supplementary Information, Fig. S2.

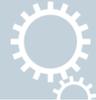

### Acknowledgments

The authors are grateful to D. I. Khomskii for helpful discussion. This work was supported by IBS-R009-D1. J.H.S. was supported by Nano·Material Technology Development Program(Green Nano Technology Development Program) through the National Research Foundation of Korea(NRF) funded by the Ministry of Education, Science and Technology (No. 2011-0030146). L.M. acknowledges support by a grant from the Swiss National Science Foundation (SNSF) (project PBELP2-125484). Y.J.C. acknowledges support from National Research Foundation of Korea under Grant No. NRF-2014R1A1A1002868. The Advanced Light Source is supported by the Director, Office of Science, Office of Basic Energy Sciences, of the U.S. Department of Energy under Contract No. DE-AC02-05CH11231. The 9C Beamline at PLS is used for this study.

### Author contributions

### Additional information

**Supplementary information** accompanies this paper at http://www.nature.com/scientificreports

**Competing financial interests**: The authors declare no competing financial interests.

**How to cite this article:** Yoo, H.K. *et al.* Latent instabilities in metallic LaNiO$_3$ films by strain control of Fermi-surface topology. *Sci. Rep.* **5**, 8746; DOI:10.1038/srep08746 (2015).